\documentclass[11pt]{article}

\usepackage{authblk}

\usepackage{graphicx}
\usepackage{dcolumn}
\usepackage{bbm}
\usepackage{bm}
\usepackage{geometry}
\usepackage{cite}
\usepackage{braket}
\usepackage{mathtools}
\usepackage{bbold}
\usepackage{subcaption}
\usepackage{amsmath,amssymb,amsthm}
\usepackage{breqn}
\usepackage{xcolor}
\definecolor{linkequation}{HTML}{1A73E8}
\usepackage[colorlinks=true,linkcolor=linkequation,citecolor=blue,urlcolor=blue]{hyperref}

\title{Directional Motional Control via Engineered Conical Intersections in Trapped Rydberg Ions}

\author[1]{Abdessamad Belfakir\thanks{Corresponding author: \href{mailto:abdobelfakir01@gmail.com}{abdobelfakir01@gmail.com}}}
\author[2]{Yousra Bouasria}
\author[3]{Herschel Rabitz}
\author[1]{Ahmed Ratnani}

\affil[1]{The UM6P Vanguard Center, Mohammed VI Polytechnic University, Rocade Rabat\mbox{-}Salé, Technopolis, 11103, Morocco.}
\affil[2]{SYRTE, Observatoire de Paris, Université PSL, CNRS, Sorbonne Université, LNE, 61 avenue de l’Observatoire, 75014 Paris, France.}
\affil[3]{Department of Chemistry, Princeton University, Princeton, NJ 08544, USA.}

\date{} 

\begin{document}
	\maketitle

	\begin{abstract}
We demonstrate coherent control of motional dynamics in trapped Rydberg ions 
engineered to exhibit a conical intersection between adiabatic potential-energy 
surfaces. Using quantum optimal control, an optimally shaped electric field 
drives the motional wave packet between prescribed 
spatial configurations on microsecond timescales. Localized nonadiabatic coupling 
breaks the symmetry of the dynamics and produces a directed trajectory: after 
only a few early passages through the conical intersection region, the packet 
proceeds toward the target with high fidelity. In contrast, in the Born–Oppenheimer 
limit, where such coupling is absent, the optimized control yields symmetric, 
multi-cycle oscillations rather than a comparably directed displacement. While 
both approaches reach the target at the chosen final time, the underlying 
trajectories are qualitatively different. This work demonstrates that engineered 
conical intersections can serve as a new control resource for directional motional 
dynamics, complementing pulse-shaping methods in trapped-ion systems. This 
directional motion of the ions, enabled by the conical intersection, is expected 
to have important applications for quantum information processing, where 
controlled motional states underpin high-fidelity gate operations and scalable 
architectures.
	\end{abstract}
	
\section{Introduction}
{Trapped ions represent one of the most promising platforms for quantum information processing ~\cite{Mueller, Bermudez, Blatt,Roos, SchmidtKaler,Kaufmann, Friis,Gulde,Riebe, Olmschenk,Figgatt,Schindler}. Achieving precise control over their motional degrees of freedom is essential for scalable quantum computation, as even small motional displacements can degrade entangling-gate performance ~\cite{PhysRevA.105.052409}. Accordingly, several protocols have been developed to suppress residual excitations including motional squeezing ~\cite{PhysRevLett.127.083201} or tailored pulse shaping ~\cite{Schafer2018}.}

{In Parallel, conical intersections (CIs), where two or more potential energy surfaces (PESs) become degenerate, are known to induce strong nonadiabatic couplings between electronic and nuclear motion ~\cite{Yarkony2, Domcke,Baer, Barbatti}. These effects play a fundamental role in ultrafast processes such as photoisomerization ~\cite{Hammarstroem}, the mechanism of vision~\cite{Schoenlein, Rinaldi,Polli}, and the photostability of DNA~\cite{Barbatti}. Recent theoretical proposals have shown that CIs can also be engineered in synthetic quantum systems~\cite{PhysRevLett.103.083201, Moiseyev_2008, Sindelka_2011}, including Rydberg atoms~\cite{Gallagher_2005, Low_2012} and trapped Rydberg ions~\cite{Gambetta}. In trapped
	Rydberg ions, where strong dipolar interactions and large electric polarizabilities enable fine-tuned control over PESs, CIs
	can be realized on micron length scales and microsecond
	timescale, enabling studies of nonadiabatic dynamics in a
	fully quantum regime with high controllability.}

{In this work, we bring these directions together by investigating the optimal control of Rydberg-ion motion near an engineered CI. Using a time-dependent electric field and shaping its temporal profile using quantum optimal control theory ~\cite{ZhuBotina,ZhuRabitz,
		Ohtsuki,Rabitz2000,Rabitz2003,Shapiro,Dantus}, we demonstrate coherent steering of a vibrational wave packet from an initial Gaussian state to a predefined spatial target. Our spinor model incorporates nonadiabatic couplings arising from the CI, allowing direct comparison with the Born–Oppenheimer (BO) limit where such couplings are absent. Near a CI, optimized fields induce a directed displacement of the wave packet with only a few early passages through the CI before reaching the target, whereas in the BO limit the control produces symmetric, multi-cycle oscillations. These results identify engineered CIs as a promising resource for realizing directed and robust motional control in trapped-ion platforms for quantum information science.}

\section{Equations of motion}
We consider two Rydberg ions of mass $m$  confined in a Paul trap, which generates an effective harmonic potential  with frequencies $\omega_x$, $\omega_y$, and $\omega_z$ \cite{Higgins2019, Higgins2019Book, Major2005}. By assuming that $\omega_y \gg \omega_x > \omega_z$, the motion of the two ions can be analyzed in the X$\text{--}$Z plane. Hence, in the position representation, the potential energy of the two trapped ions is given by
$V_{\text{trap}} = \frac{m}{2} \left[ \omega_x^2 (X_1^2 + X_2^2) + \omega_z^2 (Z_1^2 + Z_2^2) \right] + V_c,$
where \(\mathbf{R}_{i} = (X_i, Z_i)\) denotes the nuclear coordinates in the laboratory frame of the \(i\)-th ion, with \(i=1,2\). \(V_c\) is the repulsive Coulomb potential expressed as \(V_c = k\,e^{2}/|\mathbf{r}|\), where \(k\) is the Coulomb constant, \(e\) is the elementary charge, and \(\mathbf{r} = \mathbf{R}_2 - \mathbf{R}_1\). By applying a static electric field \(u_0\) along the \(X\) axis, a potential 
$V_m = u_0 \, \phi(X)$ is generated, 
where \(\phi(X) = e\,(X_1 + X_2)\) \cite{Gambetta}. The static electric field controls the transverse equilibrium positions of the ions and thus enables the construction of the CI.

Each ion can be modeled as a two-level system with \(\ket{\downarrow} \equiv \ket{n\,S} = (0,1)^T\) and \(\ket{\uparrow} \equiv \ket{n\,P} = (1,0)^T\), where \(n\) is the principal quantum number. The two Rydberg levels are coupled by a microwave  field with Rabi frequency \(\Omega\) and detuning \(\Delta\) \cite{Gambetta}.  Moreover, the two Rydberg ions interact via an exchange Hamiltonian governed by 
$ H_{ex}=U_{ex}(r)\otimes \left(\sigma_{+}^{1}\otimes\sigma_{-}^{2}+\sigma_{-}^{1}\otimes\sigma_{+}^{2}\right),
$ where $\sigma_{+}^{i}\ket{\downarrow}^{i}=\ket{\uparrow}^{i}$, $\sigma_{-}^{i}\ket{\uparrow}^{i}=\ket{\downarrow}^{i}$ and $U_{ex}(r)$ is the exchange interaction potential. Additional details on how the exchange interaction potential is engineered can be found in \cite{Gambetta}. Furthermore, due to the polarizabilities of the Rydberg states denoted by $\rho_{\sigma}$ with $\sigma=\{\downarrow,\uparrow\}$, the ions experience an additional potential given as
$H_{\rho}= -\alpha^2\left( X_{1}^2 \otimes \Pi_{\rho}\otimes{\mathbbm{1}}_{2} +X_{2}^2 \otimes {\mathbbm{1}}_{2}\otimes \Pi_{\rho} \right),$
where $\Pi_{\rho}=\rho_{\downarrow}\ket{\downarrow}\bra{\downarrow}+\rho_{\uparrow}\ket{\uparrow}\bra{\uparrow}$ and $\alpha$ is the gradient of the radio frequency of the trap \cite{Gambetta}. 

We analyze the dynamics in the center-of-mass and relative coordinate frames, defined as \( \mathbf{R} = \frac{\mathbf{R}_1 + \mathbf{R}_2}{2} \) and \( \mathbf{r} = \mathbf{R}_2 - \mathbf{R}_1 \), respectively. We restrict our analysis to the single-excitation subspace spanned by \( \ket{\pi_1} = \ket{\uparrow\downarrow} \) and \( \ket{\pi_2} = \ket{\downarrow\uparrow} \), realized by setting \( \Omega = 0 \) and holding \( \Delta \) constant~\cite{Wuster2011,Gambetta}. In this subspace, we define the effective Pauli operators as
\( S_0 = \ket{\pi_1}\bra{\pi_1} + \ket{\pi_2}\bra{\pi_2} \),
\( S_x = \ket{\pi_1}\bra{\pi_2} + \ket{\pi_2}\bra{\pi_1} \), and
\( S_z = \ket{\pi_1}\bra{\pi_1} - \ket{\pi_2}\bra{\pi_2} \).
For typical experimental conditions, the harmonic trapping potential dominates the energy landscape, allowing the remaining interactions to be treated as perturbations. We perform an harmonic expansion around the equilibrium positions \( \mathbf{R}_0 = (X_0, Z_0) \) for the center-of-mass coordinate and \( \mathbf{r}_0 = (x_0, z_0) \) for the relative coordinate, introducing displacements \( \mathbf{Q} = \mathbf{R} - \mathbf{R}_0 \) and \( \mathbf{q} = \mathbf{r} - \mathbf{r}_0 \). The derivation of the equilibrium positions \( X_0, Z_0 = 0 \) and \( x_0=0, z_0 \) is provided in Appendix A.

\begin{figure}[h]
	\centering
	\includegraphics[width=\columnwidth]{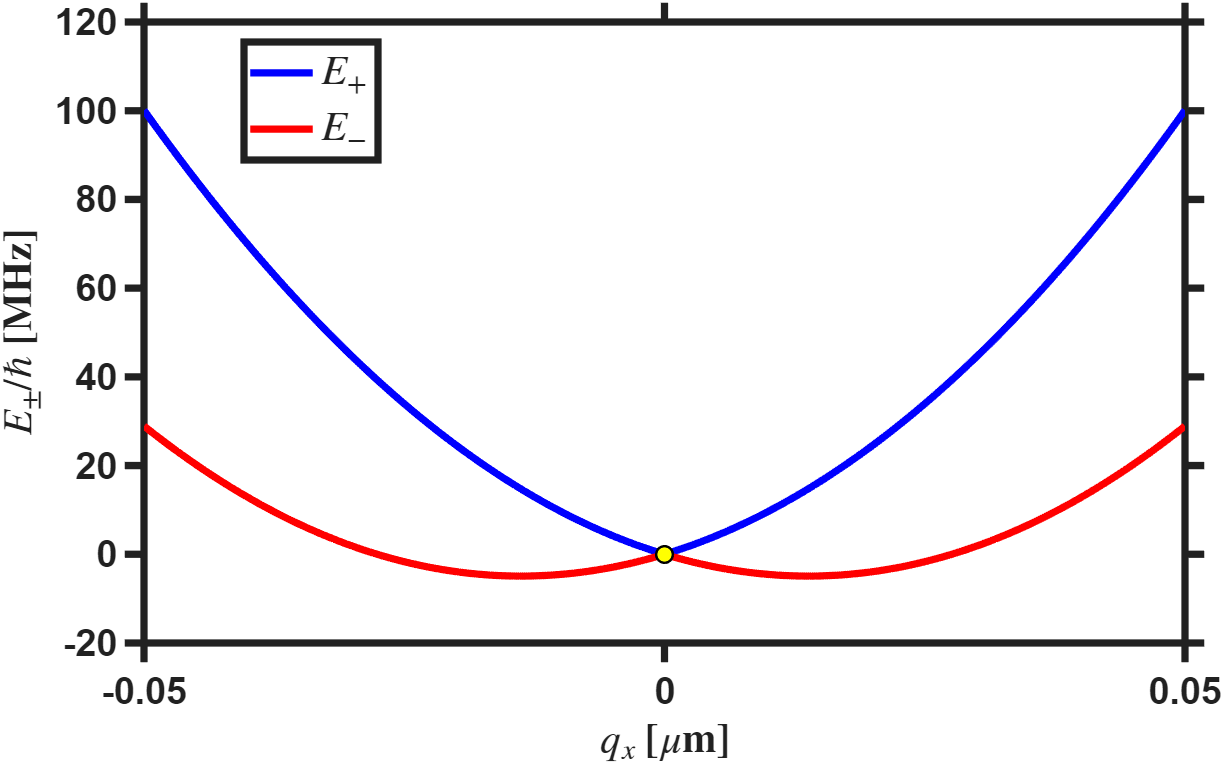}
	\caption{One-dimensional slice of the PESs at \( q_z = 0 \), showing the lower (\( E_- \), red) and upper (\( E_+ \), blue)   as functions of \( q_x \) (in \(\mu\)m). 
		The yellow dot marks the position of the CI at \( (q_x^* = 0,\, q_z^* = 0) \). 
		Parameters correspond to \( ^{88}\mathrm{Sr}^+ \) Rydberg ions: \( m = 87.9 \times 1.66 \times 10^{-27}\,\mathrm{kg} \), \( \rho_{\downarrow} = 8.9 \times 10^{-30}\,\mathrm{C}^2\mathrm{m}^2/\mathrm{J} \), and \( \rho_{\uparrow} = -3.8 \times 10^{-31}\,\mathrm{C.}^2\mathrm{m}^2/\mathrm{J} \) (Corresponding to Rydberg $\ket{nS}$ and $\ket{nP}$ states with $n=50$). Additional parameters used are \( \omega_x = 2\pi \times 1.6\,\mathrm{MHz} \), \( \omega_z = 2\pi \,\mathrm{MHz} \), \( u_0 = 2.529\,\mathrm{V/m} \) corresponding to \( X_0 = -0.024\,\mu\mathrm{m} \), \( \alpha = 8.17 \times 10^8\,\mathrm{V/m^2} \), \( U_{\mathrm{ex}}(r_0) = 0 \),  $z_0=4.31\hspace{0.05cm} \mu{m}$ and \( F_0/\,\hbar = 2\pi \times 20\,\mathrm{MHz}/\mu\mathrm{m} \).
	}
	
	\label{diagram_and_CI}
\end{figure}

Following Ref.~\cite{Gambetta}, we neglect the center-of-mass motion (\( \mathbf{Q} = 0 \)) and focus on the relative coordinate, which captures the key nonadiabatic effects. The resulting effective Hamiltonian incorporating all relevant interaction terms is given as,
\begin{subequations}\label{H_0}
	\begin{align}
		\label{eq:H0} 
		&H_0 =  -\frac{\hbar^2}{2\,\mu} \left( \nabla_{q_x}^2 + \nabla_{q_z}^2 \right)
		\otimes S_0 +H_{\text{spin}},\\
		& \text{where,}\notag\\
		&H_{\text{spin}}=S\left(\textbf{q}\right)\otimes{S}_0+W\left(\textbf{q}\right)\otimes{S}_x+G\left(\textbf{q}\right)\otimes{S}_z,
		\\
		& \text{with,}\notag\\
		&S\left(\textbf{q}\right)=\left[\frac{\mu}{2} \,\bar{\omega}_x^2\,q_x^2 + \frac{\mu}{2}\,\bar{\omega}_z^2 \,q_z^2\right],\\ &W\left(\textbf{q}\right)=\left( U_{\text{ex}}(r_0) + F_z(r_0)\,q_z \right),  \\& G\left(\textbf{q}\right) =\alpha^2\,\rho_{-} \,X_0\,q_x.
	\end{align}
\end{subequations}
Here, the reduced mass is \( \mu = m/2 \), $\hbar$ is the reduced Planck's constant and the effective  frequencies are $
\bar{\omega}_x^2 = \omega_x^2 - \frac{\alpha^2}{2\,\mu} \,\rho_+ - \frac{k\,e^2}{\mu\,|z_0|^3}, \quad
\bar{\omega}_z^2 = \omega_z^2 + \frac{2\,k\,e^2}{\mu\,|z_0|^3},$
with \( F_z(r_0) = \left. \partial_z U_{\text{ex}}(r) \right|_{r = r_0} \) and $\rho_{\pm}=\rho_{\uparrow}\pm\rho_{\downarrow}$. A detailed derivation of the effective Hamiltonian, starting from the full spinor model and including the harmonic expansion around equilibrium, is provided in Appendix~A. The PESs refer to the eigenvalues of the electronic Hamiltonian $H_{\text{spin}}$ \cite{Domcke2004,Ryabinkin2017}. These eigenvalues are defined as,
\begin{equation}\label{PES}
	E_{\pm} = S\left({\textbf{q}}\right) \pm \sqrt{G\left({\textbf{q}}\right)^2 + W\left(\textbf{q}\right)^2}.
\end{equation}
In general, CIs occur at positions \( \mathbf{q}^{*}\) where the adiabatic potentials become degenerate, i.e., \( E_+\left(\mathbf{q}^{*}\right) = E_-\left(\mathbf{q}^{*}\right) \). According to Eq.\,(\ref{PES}), this condition is equivalent to \( G\left(\mathbf{q}^{*}\right) = 0 \) and \( W\left(\mathbf{q}^{*}\right) = 0 \). This implies that the system described by Eq.~(\ref{PES}) exhibits a CI at \( q_x^{*} = 0 \) and \( q_z^{*} = -U_{\mathrm{ex}}(r_0 )/ F_z(r_0) \). Fig.\,(\ref{diagram_and_CI}) shows a one dimensional slice of the PESs at $q_z=0$ and the position of the CI for two trapped \( ^{88}\mathrm{Sr}^+ \) Rydberg ions.

In this work, we demonstrate that the motional state of two trapped Rydberg ions near a CI can be coherently controlled using a time-dependent electric field applied to a single ion. This control field, designed through quantum optimal control theory, steers the system from an initial spatial configuration to a desired target at a fixed final time. We consider a spatially localized electric field \( u(t) \) applied to the first ion along the \( X \)-axis, resulting in an interaction term \( V_u(t) = e\, u(t)\, X_1 \). Spatially selective control is experimentally feasible using tightly focused laser beams or near-field microwave radiation~\cite{PhysRevA.87.013437,PhysRevA.60.R3335, PhysRevA.60.145, Wang_2009, PhysRevLett.102.073004}. {In our work, we model the control field as a generic time-dependent electric drive, which makes the results broadly applicable independent of the specific implementation}. After performing the harmonic approximation, the control Hamiltonian can be given by,
\begin{equation}
	H_u(t) = e\, u(t) \left( X_0 - \frac{1}{2} q_x \right) \otimes S_0. \label{Control_Hamiltonia}
\end{equation}

The dynamics of the system are governed by the following Schrödinger equation,
\begin{equation}\label{Schro}
	\dfrac{\partial}{\partial t} \ket{\psi\left(q_x,q_z, t\right)} = \dfrac{-i}{ \hbar }\left(H_0 + H_u(t)\right) \ket{\psi\left(q_x,q_z, t\right)},
\end{equation}
where $\ket{\psi(q_x,q_z,t)}$ belongs to the Hilbert space \(\mathcal{H}=\mathcal{H}_q \otimes \mathcal{H}_{\mathrm{spin}},\)
with \(\mathcal{H}_q\) is the single-particle motional Hilbert space in two spatial
dimensions and \(\mathcal{H}_{\mathrm{spin}}\cong \mathbb{C}^2\) is the
two-dimensional Hilbert space associated with a spin-\(\tfrac{1}{2}\) degree of freedom.
For a known external electric field, the time-dependent Schrödinger equation {given in Eq. (\ref{Schro})} can be solved numerically. However, in quantum control problems, the form of the electric field is unknown and needs to be designed in order to minimize or maximize a cost function \cite{ZhuBotina,ZhuRabitz}. As seen in Eq.\,\eqref{Control_Hamiltonia}, the control field couples directly to \( q_x \), enabling efficient manipulation of the ions’ relative motion. If instead the control field is applied symmetrically to both ions, the resulting control Hamiltonian becomes
\( H_{u}^{B}(t) = 2\,e\,u(t)\,X_0 \otimes S_0 \). Therefore, applying this control Hamiltonian introduces only a constant energy shift and does not provide the nuanced control achievable with the control Hamiltonian given in Eq.\,\eqref{Control_Hamiltonia}. We therefore apply the time-dependent electric field to a single ion to achieve full motional control.

\section{Control of motional states of trapped Rydberg ions}
Our objective is to determine the optimal electric field \( u(t) \) that drives the system from an initial to a target state at a fixed final time \( t_f \). The initial state is
\(|\psi\left(q_x,q_z,0\right)\rangle=\phi^0\left(q_x,q_z\right)|\pi_2\rangle\),
where
$
\phi^0\left(q_x,q_z\right)=\ell_{xz}\exp\!\left[-\frac{\mu}{2\,\hbar}\big(\bar\omega_x(q_x-q_x^0)^2+\bar\omega_z(q_z-q_z^0)^2\big)\right],
$ with 
$\ell_{xz}=\left(\frac{\mu^2\bar\omega_x\bar\omega_z}{\pi^2\hbar^2}\right)^{\!1/4}.
$
The target motional state 
\(\phi^d(q_x,q_z)\) is a Gaussian with the same widths as
\(\phi^0(q_x,q_z)\) but centered at \((q_x^d,q_z^d)\).  We define the target operator  as
\begin{equation}
	\label{targetfunction}
	\hat{O} = \Pi_{\ket{\phi^d}\bra{\phi^d}} \otimes S_0,
\end{equation}
where the action of the projector $\Pi_{\ket{\phi^d}\bra{\phi^d}}$ on a wave function $\Psi\left(q_x, q_z\right)\in\mathcal{H}_q$ is given by
$
\left( \Pi_{\ket{\phi^d}\bra{\phi^d}} \Psi \right)(q_x, q_z) 
= \phi^d(q_x, q_z) 
\quad \times \int \int dq_x' dq_z' \, (\phi^d(q_x', q_z'))^* \Psi(q_x', q_z').
$
Our goal is to maximize the expectation value of the projector \( \hat{O} \)
while minimizing the energy of the control field, subject to the Schrödinger dynamics given in Eq.~(\ref{Schro}). Following \cite{ZhuRabitz,ZhuBotina}, this objective can be equivalently expressed as the maximization of the following functional,
\begin{equation}\label{functional}
	J = J_1 - J_2,    
\end{equation}
where
\begin{subequations}\label{eq:J_functional}
	\begin{align}
		J_1 &= \braket{\psi\left(q_x,q_z,t_f\right)|\hat{O}|\psi\left(q_x,q_z,t_f\right)} 
		\notag \\&- 2\,\Re \int_0^{t_f} dt\, \bra{\lambda\left(q_x,q_z,t\right)}\left( \frac{\partial}{\partial t} - \hat{L} \right)\ket{\psi\left(q_x,q_z,t\right)}, \label{eq:J1} \\
		J_2 &= \alpha_0 \int_0^{t_f} \left[u(t)\right]^2\, dt. \label{eq:J2}
	\end{align}
\end{subequations}
The first term in \( J_1 \) represents the expectation value of \( \hat{O} \), while the second term ensures the correct physical evolution of the system. \( J_2 \) imposes a constraint on the control field energy. Here, \( \ket{\lambda(q_x,q_z,t)} \) is a Lagrange multiplier used to enforce the system's evolution, and \( \hat{L} \) is the operator given on the right-hand side of Eq.\,\eqref{Schro}. The parameter \( \alpha_0 \) is a positive weighting factor that controls the contribution of the field fluence. In order to derive the optimal control capable of transitioning the system from the initial state to the target state, we apply the  monotonically convergent algorithm (MCA) summarized in Appendix~B. The MCA iteratively updates the control field to maximize the objective functional while ensuring stable convergence. Before starting the optimization, an initial guess for the control field is specified. Then, the control field is refined over successive iterations of the algorithm until the objective functional reaches its maximum.

In our simulations, the initial wave packet center is set to \( (q_x^0, q_z^0) = (-11.4\,\mathrm{nm},\,0) \), the target wave packet center to \( (q_x^d, q_z^d) = (11.4\,\mathrm{nm},\,0) \), and the final time to \( t_f = 10\,\mu\mathrm{s} \).
For the optimization parameters listed in Appendix~B, we take \(\eta=\zeta=1\).
The fluence penalty in Eq.\,\eqref{eq:J2} is \(\alpha_0=0.01\).
The initial electric field used to seed the optimization is a rectangular pulse,
\(
\mathcal{E}\left(t\right)  
=-14.39~\mathrm{V.m}^{-1} \) for $t \in [0, 20 ]\,\mathrm{ns} $ and  $0$ otherwise.

\begin{figure*}[t]
	\centering
	\includegraphics[width=\textwidth]{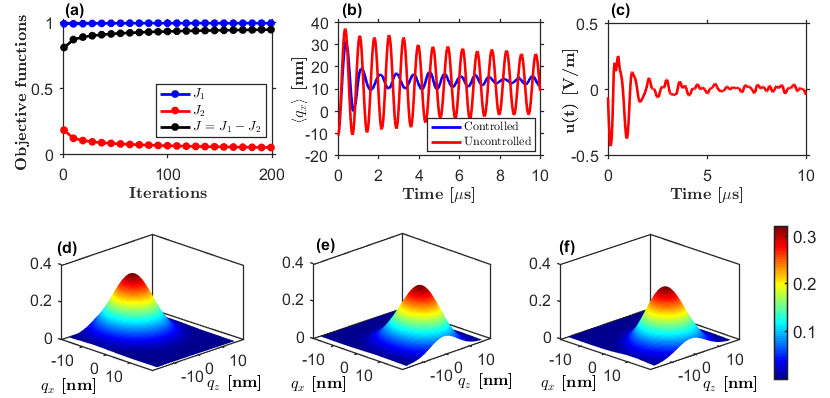}
	\caption{Optimal control of spatial dynamics near a CI.
		(a) Convergence of the target functional \(J_1\), the penalty \(J_2\), and the total \(J = J_1 -  J_2\) over the MCA iterations.
		(b) Time evolution of \(\langle q_x\left(t\right)\rangle\) with (blue) and without (red) the control field, demonstrating successful steering.
		(c) Temporal profile of the optimized electric field $u(t)$.
		(d--f) Probability density \(|\psi\left(q_x, q_z, t\right)|^2\) at (d) \(t=0~\mu\text{s}\), (e) \(t=5~\mu\text{s}\), and (f) \(t=10~\mu\text{s}\), illustrating transport to the target region under control.
		Additional parameters are provided in the main text and in Fig.\,\ref{diagram_and_CI}.}
	
	\label{Control_position_exact}
\end{figure*}

In Fig.\,\ref{Control_position_exact}(a), we show the target function \( J_1 \), the penalty term \( J_2 \), and the total functional \( J = J_1 - J_2 \) as a function of the algorithm iteration number. The functional \( J \) increases rapidly in the first 20 iterations, then stabilizes after approximately 100 iterations. At the final iteration, \( J_1 \) reaches 99.08\%, indicating a high overlap with the target operator \( \hat{O} \), while \( J_2 \) decreases. To confirm the control objective, Fig.\,\ref{Control_position_exact}(b) shows the time evolution of the position operator \( q_x \) with and without the optimized electric field $u(t)$. The relative position between the ions exhibits oscillatory behavior, and under the action of the optimal field, it reaches the desired value at the final time, indicating successful spatial control. { By contrast, in the absence of the control field the final expectation value is 
	\(\langle q_x(t_f)\rangle = 8.96~\mathrm{nm}\), compared to the desired 
	\(q_x^d = 11.4~\mathrm{nm}\)}. From Fig.~\ref{Control_position_exact}(b), one can conclude also that the optimized field induces a directed displacement toward the target: the wave packet traverses the vicinity of the CI point twice at early times and does not return thereafter.

{Figs.\,\ref{Control_position_exact}(d), \ref{Control_position_exact}(e), and \ref{Control_position_exact}(f) show the density \( |\psi(q_x, q_z, t)|^2 \) at three representative times: initial, intermediate, and final. These snapshots illustrate the real-space dynamics of the wavefunction under the optimized field, and demonstrate that the control objective has been achieved by driving the system from the initial to the target configuration.}

\begin{figure*}[t]
	\centering
	\includegraphics[width=\textwidth]{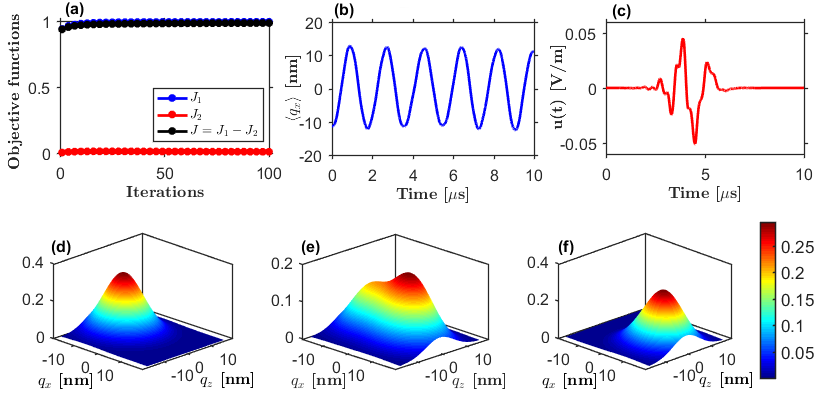}
	\caption{
		Optimal control of wave packet dynamics under the  BO approximation. 
		(a) Evolution of the target functional \( J_1 \), penalty \( J_2 \), and total \( J = J_1 - J_2 \) over the MCA iterations. 
		(b) Time evolution of \( \langle q_x\left(t\right) \rangle \) under optimal control. 
		(c) Time profile of the optimized electric field $u(t)$. 
		(d–f) Density \( |\psi\left(q_x, q_z, t\right)|^2 \) at representative times: (d) \( t = 0~\mu\text{s} \), (e) \( t = 5~\mu\text{s} \), and (f) \( t = 10~\mu\text{s} \). Additional parameters are provided in the main text and in Fig.\,\ref{diagram_and_CI}.
	}
	\label{Control_position_BO}
\end{figure*}

\section{Control Dynamics in the absence of the CI (BO Limit)}
To isolate the effects of the CI on the control of the motion dynamics, we solve the same optimal control problem within the BO  approximation. This approximation confines the system’s dynamics to a single adiabatic PES, neglecting nonadiabatic couplings between electronic states. Under this approach, the full spinor structure of the Hamiltonian is removed, and the dynamics is governed by the time-dependent Schrödinger equation with the Hamiltonian,
$
H^{\mathrm{BO}} = -\frac{\hbar^2}{2\,\mu} \nabla_{q_x}^2 - \frac{\hbar^2}{2\,\mu} \nabla_{q_z}^2 + E_{-} + e\, u(t) \left( X_0 - \frac{1}{2} q_x \right),
$
where $E_{-}$ is given in Eq.\,(\ref{PES}).
We apply the same optimization procedure to the reduced model, targeting the same objective operator \( \hat{O} \).  The optimization parameters, shown in Appendix~B, are \(\eta=\zeta=1\) and the fluence penalty \(\alpha_{0}=0.1\).
The initial electric field used before optimization is a simple Gaussian-shaped pulse along the $X$-direction, defined as :
$
\mathcal{E}\left(t\right) = \mathcal{E}_0 \cdot \exp\left[ -\frac{(t - t_0)^2}{2\,\sigma^2} \right] $
where the pulse center \( t_0 = 4~\mu\mathrm{s} \), \( \mathcal{E}_0 = -0.719~\mathrm{mV/m} \) and \( \sigma =   0.85~\mu\mathrm{s} \).
This field served as the initial guess for the optimization algorithm.

In Fig.~\ref{Control_position_BO}(a), we show the evolution of the target function \( J_1 \), the penalty term \( J_2 \), and the total functional \( J = J_1 - J_2 \) as a function of the algorithm iteration number for the BO limit. By comparing the evolution of the objective functional $J$ in the BO case with that in the full spinor dynamics case (including the CI) shown in Fig.\,\ref{Control_position_exact}(a), we observe that in the   BO case, the total functional $J$ converges to its optimal value in significantly fewer iterations—approximately 30 compared to 100 for the exact dynamics. This indicates that the CI complicates the control process due to the presence of nonadiabatic couplings. 

\begin{figure}[h]
	\centering
	\includegraphics[width=\columnwidth]{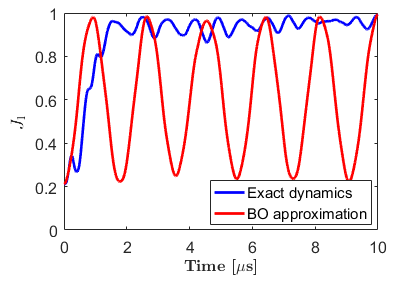}
	\caption{Time evolution of $J_1$.}
	\label{J_1_figure}
\end{figure}

In Fig.\,\ref{Control_position_BO}(b), we show the time evolution of the expectation value \( \langle q_x\left(t\right) \rangle \) under optimal control for  the BO approximation.  {The control field produces regular oscillations, such that 
	the wave packet repeatedly traverses the spatial region that corresponds to 
	the CI point in the full spinor model. The absence of nonadiabatic coupling 
	preserves symmetry in the motion, resulting in periodic dynamics}. In contrast, as seen in Fig.~\ref{Control_position_exact}(b), the  dynamics exhibits two early passages through the vicinity of the CI point, after which the wave packet proceeds toward the target without further return. This contrast identifies the localized nonadiabatic coupling near the engineered CI—together with optimal field shaping—as the mechanism enabling directional motional control. { It is known that even a small amount of displacement of trapped ions erodes 
	the performance of entangling gates~\cite{PhysRevA.105.052409}, while 
	protocols based on motional squeezing have been shown to enable clean, 
	residual-free transport and separation of ions~\cite{PhysRevLett.127.083201}. 
	Moreover, tailored pulse shaping has been successfully employed to design 
	robust motional trajectories for high-fidelity gate operations in trapped-ion 
	experiments~\cite{Schafer2018}}. In this context, the directional movement of 
the relative ions induced by the presence of the CI can provide a useful resource for quantum information science.
{Fig.~\ref{J_1_figure} compares the time evolution of the target 
	functional \(J_{1}\) for the exact dynamics including the CI and for the 
	BO approximation. With the CI, \(J_{1}\) rapidly approaches unity and 
	remains stable, whereas in the BO limit it exhibits large oscillations, 
	signaling repeated departures from the target. Directed motion in 
	trapped-ion systems has also been demonstrated in the transport of 
	topological solitons in Coulomb crystals~\cite{PhysRevLett.119.153602}. In analogy, the engineered CI provides a complementary route to convert 
	oscillatory dynamics into robust directed transport, a feature that is 
	promising for enhancing coherence and stability in future trapped-ion 
	applications.
}

In Fig.\,\ref{Control_position_BO}(c), we present the optimized electric field  $u(t)$ under the BO approximation. Compared to the full dynamics with the CI, the control energy is significantly lower. This reduction reflects the simplified optimization landscape in the absence of nonadiabatic couplings. As a result, the optimization algorithm requires a less intricate field to reach the target, reducing the field amplitude and total fluence needed for optimal control.

{Figs.\,\ref{Control_position_BO}(d–f) display the wavefunction density 
	\( |\psi(q_x, q_z, t)|^2 \) at representative times under the BO 
	approximation, showing the evolution from the initial to the target 
	configuration. 
}

\section{Conclusions}
{
	We have demonstrated coherent control of motional dynamics in trapped Rydberg ions near a CI using an optimally shaped time-dependent electric field. To isolate the role of nonadiabatic effects, we compared the full spinor dynamics with the BO approximation. With the CI present, the optimized field drives the wave packet toward the target on microsecond timescales, producing a directed motional displacement. In the BO limit (no CI), the control produces symmetric, multi-cycle oscillations of large magnitude rather than a comparably directed displacement within the same time interval. This contrast identifies the localized nonadiabatic coupling near the engineered CI—together with optimal field shaping—as the mechanism enabling directional motional control. While both cases reach high fidelity at the chosen final time, the CI case achieves it through a qualitatively different trajectory.  { This prospect  opens up a pathway toward overcoming one of the central limitations in trapped-ion quantum computing—residual motional excursions that degrade gate fidelity. By adding CIs to the trapped-ion control toolbox, this work lays the groundwork for implementing future strategies that could significantly advance the fidelity and scalability of ion-based quantum information platforms.}
 
\section*{Acknowledgements:} A.B. acknowledges his colleagues at the UM6P Vanguard Center for valuable discussions on control theory and for their insightful feedback and support. H.R. acknowledges support from the U.S. Department of Energy under Grant No. DE-FG02-02ER15344.

\section*{Appendices}

\appendix
\renewcommand{\theequation}{A.\arabic{equation}}
\setcounter{equation}{0} 
 
{\section*{Appendices}
	\appendix
	\section{Appendix A: Derivation of the Effective Hamiltonian}\label{Appdix A}
	\renewcommand{\theequation}{A.\arabic{equation}}
	\setcounter{equation}{0}
	We provide here the derivation of the effective Hamiltonian used in the main text. The total Hamiltonian includes the trap potential, Coulomb repulsion, exchange interaction, polarizability-induced potential, the microwave coupling and the external control field. The full Hamiltonian is given by
	\begin{equation}
		H=H_0+H_u(t),
	\end{equation}
	where
	\begin{equation}\label{Full_Hamiltonian}
		H_0= \left[ T_{\text{n}} + V_{\text{trap}} + V_{m}\right] \otimes \mathbbm{1}_2 \otimes \mathbbm{1}_2 
		+ H_{\text{ex}} +{\mathbbm{1}}_{n}\otimes{H}_{MW}+ H_{\rho},
	\end{equation}
	and \begin{equation}
		H_u(t)= V_u(t) \otimes \mathbbm{1}_2 \otimes \mathbbm{1}_2
	\end{equation}
	where $V_{\text{trap}}$, $V_{m}$, $H_{ex}$, $H_{\rho}$ and $V_{u}(t)$ are defined in the main text,  ${\mathbbm{1}}_{n}$ is the nuclear identity operator, and $T_n$ is the kinetic energy operator expressed as 
	\begin{equation}
		T_n=-\dfrac{\hbar^2}{2\,m}\nabla_{\textbf{R}_1}^2-\dfrac{\hbar^2}{2\,m}\nabla_{\textbf{R}_2}^2.
	\end{equation}
	The two Rydberg levels are coupled by a microwave  field with Rabi frequency \(\Omega\) and detuning \(\Delta\) \cite{Gambetta}. This coupling Hamiltonian appearing in Eq.\,(\ref{Full_Hamiltonian}) can be described by \begin{equation}\label{HMW}
		H_{\text{MW}}=\hbar\,\Omega \,\left(\sigma_{x}^{1}\otimes{\mathbbm{1}}_{2}+{\mathbbm{1}}_{2}\otimes\sigma_{x}^{2}\right) + \hbar\,\Delta\,\left(\sigma_{z}^{1}\otimes{\mathbbm{1}}_{2}+{\mathbbm{1}}_{2}\otimes\sigma_{z}^{2}\right).
	\end{equation} 
	We now recast this Hamiltonian in center-of-mass and relative coordinates: $
	\mathbf{R} = \frac{\mathbf{R}_1 + \mathbf{R}_2}{2},  \mathbf{r} = \mathbf{R}_2 - \mathbf{R}_1$. 
	This leads to 
	\begin{equation}\label{H0}
		H_{0}=H_{CoM}+H_{rel}+H_{CoM-rel},
	\end{equation}
	where $H_{CoM}$ depends solely on the centre of mass coordinates, $H_{rel}$ depends on relative coordinates, while $H_{CoM-rel}$ depends on both sets of coordinates \cite{Gambetta}. The expressions  of these Hamiltonians are as follows
		\begin{equation}
			H_{CoM}=\bigg[-\dfrac{\hbar^2}{2M}\left(\nabla_{X}^2+\nabla_{Z}^2\right)+\dfrac{M}{2}\Big(\omega_{x}^{2}X^{2}+\omega_{z}^{2}Z^{2}\Big)+2\,{u_{0}}\,e\,{X}\bigg]
			\otimes {\mathbbm{1}}_{2}\otimes {\mathbbm{1}}_{2}-\alpha^2 X^2\otimes \bigg[\Pi_{\rho}\otimes{\mathbbm{1}}_{2}+{\mathbbm{1}}_{2}\otimes\Pi_{\rho}\bigg],
		\end{equation}
		\begin{equation}
			H_{rel}=\bigg[-\dfrac{\hbar^2}{2\mu}\left(\nabla_{x}^2+\nabla_{z}^2\right)+\dfrac{\mu}{2}\Big(\omega_{x}^{2}x^{2} +\omega_{z}^{2}z^{2}\Big)+V_{c}\bigg] \otimes {\mathbbm{1}}_{2}\otimes {\mathbbm{1}}_{2} -\alpha^2\,\dfrac{x^2}{4}\otimes \bigg[\Pi_{\rho}\otimes{\mathbbm{1}}_{2}+{\mathbbm{1}}_{2}\otimes\Pi_{\rho}\bigg]+H_{ex}+{\mathbbm{1}}_{n}\otimes H_{\text{MW}},
		\end{equation}
	and
	\begin{equation}
		H_{CoM-rel}= \alpha^2\,{X}\,{x}\otimes\bigg[\Pi_{\rho}\otimes{\mathbbm{1}}_{2}-{\mathbbm{1}}_{2}\otimes\Pi_{\rho}\bigg],
	\end{equation}
	where $\Pi_{\rho}$ is defined in the main text and $M=2m$.
	In the following, we restrict our analysis to the case where only one ion is excited to the  Rydberg state $\ket{\uparrow}$, while the second ion is prepared in the Rydberg level $\ket{\downarrow}$. This condition requires that $\Omega=0$ and $\Delta$ remains constant \cite{Wuster2011, Gambetta}. Hence, the dynamics can be fully described within the Hilbert space  spanned by the vectors $\ket{\pi_{1}}=\ket{\uparrow\downarrow}$ and $\ket{\pi_{2}}=\ket{\downarrow\uparrow}$. 
	it follows that 
	\begin{dmath}
		H_{CoM} = \bigg[-\dfrac{\hbar^2}{2M}\left(\nabla_{X}^2+\nabla_{Z}^2\right) +\dfrac{M}{2}\Big(\omega_{x}^{2}\,X^{2} + \omega_{z}^{2}\,Z^{2}\Big) -\alpha^2\,X^2\, \rho_{+} + 2\,{u_{0}}\,{e}\,{X}\bigg] \otimes S_{0},
	\end{dmath}
	\begin{dmath}
		H_{rel}=\bigg[-\dfrac{\hbar^2}{2\,\mu}\left(\nabla_{x}^2+\nabla_{z}^2\right)+\dfrac{\mu}{2}\Big(\omega_{x}^{2}\,x^{2}+\omega_{z}^{2}\,z^{2}\Big) +\dfrac{k\,{e}^{2}}{\left(x^2+z^2 \right)^{1/2}} -\alpha^2\rho_{+}\,\dfrac{x^2}{4}\bigg] \otimes {S_0}+U_{ex}(x,z)\otimes {S_{x}},
	\end{dmath}
	and 
	\begin{equation}
		H_{CoM-rel}= \alpha^2\,\rho_{-}\,{X}\,{x}\otimes{S}_{z},
	\end{equation}
	where $\rho_{\pm}=\rho_{\uparrow}\pm\rho_{\downarrow}$ and $S_0$, $S_x$ and $S_z$ are defined in the main text.
	Thus, the Hamiltonian $H_{0}$ of Eq.\,\eqref{H0} can be expressed in the following form 
	\begin{equation}
		H_{0}=\bigg[-\dfrac{\hbar^2}{2\,M}\left(\nabla_{X}^2+\nabla_{Z}^2 \right)-\dfrac{\hbar^2}{2\,\mu}\left(\nabla_{x}^2+\nabla_{z}^2 \right)\bigg]\otimes{S}_{0}+H_{\text{spin}},
	\end{equation}
	where the Hamiltonian $H_{\text{spin}}$ is given by 
	\begin{equation}\label{Hspin}
		H_{\text{spin}}=\left(V_{CoM} +V_{rel}\right)\otimes S_{0}+U_{ex}(x,z)\otimes {S_{x}}+H_{CoM-rel}
	\end{equation}
	with
	\begin{equation}
		V_{CoM}=\dfrac{M}{2}\left(\omega_{x}^{2}\,X^{2}+\omega_{z}^{2}\,Z^{2}\right)-\alpha^2\, X^2 \,\rho_{+}+2\,{u_{0}}\,{e}\,X,
	\end{equation}
	and 
	\begin{equation}
		V_{rel}=\dfrac{\mu}{2}\left(\omega_{x}^{2}\,x^{2}+\omega_{z}^{2}\,z^{2}\right) +\dfrac{k\,{e}^{2}}{\sqrt{x^2+z^2}}-\alpha^2\,\rho_{+}\,\dfrac{x^2}{4}.
	\end{equation}
	The control Hamiltonian becomes,
	\begin{equation}\label{control_Hamiltonian}
		H_u\left(t\right)=V_u\left(t\right) \otimes {S}_0.
	\end{equation}
	For particular values of the system parameters, the terms $U_{ex}\left(x,z\right)\otimes {S_{x}}$, $H_{CoM-rel}$, and $H_{u}\left(t\right)$ can be treated as small perturbations to $\left(V_{CoM} +V_{rel}\right)\otimes S_{0}$ \cite{Gambetta}. Hence, the equilibrium positions $R_{0}$ and $r_{0}$ of the ions can be determined by solving
	$ 
	\dfrac{\partial V_{CoM}}{\partial X}\Big|_{X_{0},Z_{0}}=0, \quad \dfrac{\partial V_{CoM}}{\partial Z}\Big|_{X_{0},Z_{0}}=0,
	$ 
	and $
	\dfrac{\partial V_{rel}}{\partial x}\Big|_{x_{0}, z_{0}}=0, \quad \dfrac{\partial V_{rel}}{\partial z}\Big|_{x_{0}, z_{0}}=0$. 
	As suggested in \cite{Gambetta}, we choose  $x_0=0$ and $z_0$ can be numerically determined from $\dfrac{\partial V_{rel}}{\partial z}\Big|_{x_{0}, z_{0}}=0$.
	By performing a harmonic approximation of $V_{CoM}$ and $V_{rel}$ around the corresponding ions’ equilibrium positions, we find
	\begin{equation}
		V_{CoM}\approx\left[\dfrac{M}{2}\,\omega_x^2-\alpha^2\,\rho_{+}\right]Q_X^2+\dfrac{M}{2}\,\omega_z^2\,Q_{Z}^2,
	\end{equation}
	and 
	\begin{equation}
		V_{rel}\approx\left[\dfrac{\mu}{2}\omega_x^2-\dfrac{\alpha^2}{4}\rho_{+}-\dfrac{ke^2}{2|z_0|^3}\right]q_x^2 + \left[\dfrac{\mu}{2}\omega_z^2+\dfrac{ke^2}{|z_0|^3}\right]q_z^2,
	\end{equation}
	where
	\[
	Q = R - R_{0} = \left(X - X_0,\, Z - Z_0 \right), \quad q = r - r_{0} = \left(x - x_0,\, z - z_0\right).
	\]
	
	Therefore, Eq.\,\eqref{Hspin} can be written in terms of operators $Q$ and $q$ as
	\begin{equation}\label{H_spin}
		H_{\text{spin}}\approx S({Q,q})\otimes S_0 + W\left(Q,q\right)\otimes S_x + G\left(Q,q\right)\otimes S_z,
	\end{equation}
	where 
	\begin{align}
		S({Q,q}) &= \left[\left(\dfrac{M}{2}\omega_x^2 - \alpha^2\rho_+\right) Q_X^2 + \dfrac{M}{2}\omega_z^2 Q_Z^2 \right. \\
		&\quad\left. + \left(\dfrac{\mu}{2}\omega_x^2 - \dfrac{\alpha^2}{4}\rho_+ - \dfrac{ke^2}{2|z_0|^3}\right) q_x^2 + \left(\dfrac{\mu}{2}\omega_z^2 + \dfrac{ke^2}{|z_0|^3}\right) q_z^2 \right], \nonumber \\
		W\left(Q,q\right) &= U_{ex}(r_0) + F_z(r_0)\,q_z, \\
		G\left(Q,q\right) &= \alpha^2 \rho_- \left(Q_X + X_0\right)\, q_x,
	\end{align}
	and we have expanded $U_{ex}(r)$ around $r_0$ with
	\[
	F_z(r_0) = \nabla_r U_{ex}(r) \Big|_{r = r_0}.
	\]
	
	The PESs  refer to the eigenvalues of the electronic Hamiltonian $H_{\text{spin}}$ \cite{Domcke2004,Ryabinkin2017}. These eigenvalues are defined as
	\begin{equation}
		E_{\pm} = S({Q,q}) \pm \sqrt{G\left(Q,q\right)^2 + W\left(Q,q\right)^2}.
	\end{equation}
	The corresponding eigenvectors are expressed as:
	\begin{equation}\label{phi_plus}
		\ket{\varphi_{+}\left(Q,q\right)} = \cos\left(\Lambda(Q,q)\right)\ket{\pi_{1}} + \sin\left(\Lambda(Q,q\right))\ket{\pi_{2}},
	\end{equation}
	and
	\begin{equation}\label{phi_moins}
		\ket{\varphi_{-}\left(Q,q\right)} = -\sin\left(\Lambda(Q,q)\right)\ket{\pi_{1}} + \cos\left(\Lambda(Q,q)\right)\ket{\pi_{2}},
	\end{equation}
	where $\Lambda(Q,q)$ is defined by the relation $\tan\left(2\,\Lambda(Q,q)\right) = W(Q,q)/G(Q,q)$.\\
	The control Hamiltonian provided in {Eq.\,(\ref{control_Hamiltonian})} can be written in terms of $Q$ and $q$ as
	\begin{equation}
		H_u(t) = e\, u(t) \left(Q_X + X_0 - \dfrac{1}{2} q_x \right) \otimes S_0.
	\end{equation}
	As shown in the Supplemental Material of Ref.~\cite{Gambetta}, the dominant contributions to the dynamics arise in the relative coordinate sector. By neglecting the center-of-mass motion (i.e., setting \( \mathbf{Q} = 0 \)) and focusing exclusively on the relative coordinate \( \mathbf{q} \), we obtain the forms of the Hamiltonian \( H_0 \) and the control Hamiltonian \( H_u(t) \) given in Eqs.\,(\ref{eq:H0}) and~(\ref{Control_Hamiltonia}), respectively.

	\renewcommand{\theequation}{B.\arabic{equation}}
	\setcounter{equation}{0} 
	\section*{Appendix B: Monotonically Convergent Optimization Algorithm}\label{Appdix B}
	
	To solve the control problem defined by the objective functional in Eq.\,(\ref{functional}), we begin by choosing an initial electric field \( u^0(t) \). The state vector is then propagated forward in time over the interval \( [0, t_f] \) using
	\begin{equation}
		\frac{\partial}{\partial t} \ket{\psi^0(t)} = \left( A + i N u^0(t) \right) \ket{\psi^0(t)}, \quad \ket{\psi^0(0)} = \ket{\psi_0},
	\end{equation}
	where the operators \( A \) and \( N \) are defined by Eq.\,(\ref{Schro}), and \( \ket{\psi^0\left(t\right)} \equiv \ket{\psi^0\left(q_x, q_z, t\right)} \). The index \( 0 \) indicates the initial iteration~\cite{Ohtsuki}.
	
	For subsequent iterations \( k \geq 1 \), the Lagrange multiplier \( \ket{\lambda^k(t)} \equiv\ket{\lambda^k(q_x,q_z,t)}\) is propagated backward in time from \( t = t_f \) to \( t = 0 \), starting from the final condition \( \ket{\lambda^k(t_f)} = \hat{O} \ket{\psi^{k-1}(t_f)} \) with \( \ket{\psi^k\left(t\right)} \equiv \ket{\psi^k\left(q_x, q_z, t\right)} \). The backward evolution is given by
	\begin{equation}
		\frac{\partial}{\partial t} \ket{\lambda^k(t)} = \left( -A^\dagger + i N^\dagger \bar{u}^k(t) \right) \ket{\lambda^k(t)},
	\end{equation}
	where
	\begin{equation} \label{u_k_bar_def}
		\bar{u}^k(t) = (1 - \eta) u^{k-1}(t) - \frac{\eta}{\alpha_0} \, \Im \left[ \braket{\lambda^k(t) | N | \psi^{k-1}(t)} \right].
	\end{equation}
	
	Next, the state vector is propagated forward again via
	\begin{equation}
		\frac{\partial}{\partial t} \ket{\psi^k(t)} = \left( A + i N u^k(t) \right) \ket{\psi^k(t)}, \quad \ket{\psi^k(0)} = \ket{\psi_0},
	\end{equation}
	where the updated control field is given by
	\begin{equation} \label{u_k_update}
		u^k(t) = (1 - \zeta) \bar{u}^k(t) - \frac{\zeta}{\alpha_0} \, \Im \left[ \braket{\lambda^k(t) | N | \psi^k(t)} \right].
	\end{equation}
	
	At each iteration \( k \), the states \( \ket{\psi^k(t)} \), \( \ket{\lambda^k(t)} \), and the control field \( u^k(t) \) are used to evaluate the objective functional \( J \). The algorithm continues until convergence is reached.

\newpage{\pagestyle{empty}\cleardoublepage}
	
\end{document}